\documentclass[showpacs,preprintnumbers,amsmath,amssymb]{revtex4}
\newtheorem{theorem}{Theorem}
\newtheorem{definition}{Definition}
\newtheorem{lemma}{Lemma}
\newtheorem{corollary}{Corollary}

\usepackage{graphicx}
\usepackage{dcolumn}
\usepackage{bm}

\begin{document}
\title{
Quantum Kolmogorov Complexity and Quantum Key Distribution
}

\author{Takayuki Miyadera}
\email{miyadera-takayuki@aist.go.jp}
\author{Hideki Imai}%
 \altaffiliation[Also at ]{Graduate School of Science and Engineering,
Chuo University.
1-13-27 Kasuga, Bunkyo-ku, Tokyo 112-8551, Japan .
 }%
\affiliation{%
Research Center for Information Security (RCIS), \\
National Institute of Advanced Industrial
Science and Technology (AIST). \\
Daibiru building 1003,
Sotokanda, Chiyoda-ku, Tokyo, 101-0021, Japan.
}%


\date{\today}

\begin{abstract}
We discuss the Bennett-Brassard 1984 (BB84)
 quantum key distribution protocol in 
the light of quantum algorithmic information. 
While Shannon's information theory needs a probability to 
define a notion of information, algorithmic information 
theory does not need it and 
can assign 
a notion of information to an individual object.
The program length necessary to describe an object, Kolmogorov 
complexity, plays the most fundamental role in the theory.
In the context of algorithmic information theory, 
we formulate a security criterion for the quantum 
key distribution by using the quantum Kolmogorov complexity 
that was recently defined by Vit\'anyi. 
We show that a simple BB84 protocol indeed 
distribute a binary sequence between Alice and Bob 
that looks almost random for Eve with a probability 
exponentially close to $1$. 
\end{abstract}
\maketitle
\section{Introduction}
Cryptography is one of the most important arts in 
modern society. It enables us to communicate 
securely with our friends who live far away. 
In 1984, Bennett and Brassard \cite{BB84} 
proposed a simple but also an astonishing 
protocol which is called the BB84 protocol.
The protocol uses quantum theory 
as its essential part
in
achieving unconditionally
secure key distribution
\cite{Mayers, Lo, Shor, Boykin}.
The security notion of the BB84 protocol is based on 
Shannon's information theory \cite{note}. 
Roughly speaking, the security criterion demands that 
a random variable representing a final key 
and another random variable representing Eve's guess are 
almost independent. That is, the Shannon entropy of the final key 
from Eve's viewpoint should attain a value sufficiently close to 
its maximum value. 
In this paper, we give an alternative point of view on this 
problem. We reconsider the protocol in algorithmic information 
theory. 
In the middle of 1960's, Kolmogorov \cite{Kolmogorov}
and independently Chaitin \cite{Chaitin} described an 
innovative idea that makes a bridge between information theory and 
computation theory.
While Shannon's conventional information theory treats 
probability distributions and needs them to define 
a notion of information, 
their theory, algorithmic information theory,
takes randomness with respect to the algorithm as the 
heart of the information. 
Their formalism thus does not need a probability to 
define information, and can assign 
a notion of information to each individual object
such as a binary sequence.
The theory has been applied to problems in various fields 
including physics \cite{LiVitanyi}. 
As entropy does in Shannon's information theory, 
in algorithmic 
information theory a quantity called 
the Kolmogorov complexity 
plays the most fundamental role. 
The Kolmogorov complexity is defined as 
the length of the shortest description of an object. 
Kolmogorov complexity has some 
good properties and behaves rather rationally, 
as does entropy in
Shannon's information theory. 
Thus, 
the security criterion that we 
are to consider in this paper should not be based on 
Shannon's entropy, but on Kolmogorov complexity instead. 
Moreover, since Eve has a quantum state, the Kolmogorov complexity 
has to be extended to be able to 
treat quantum states as its 
inputs. 
That is, a secure final key should have sufficiently 
large quantum Kolmogorov 
complexity for Eve.
\par
Recently, some versions of quantum Kolmogorov complexity 
have been proposed. We employ one of them 
which was defined by Vit\'anyi \cite{Vitanyi}. It has a natural 
interpretation in terms of classical programs 
for quantum Turing machines.
In Sec. \ref{sec:QKC}, we give a brief review of Vit\'anyi's definition. 
Its two properties 
that play important roles in our paper are explained. 
In Sec. \ref{subsec:OTP}, we discuss 
the security
that can be attained 
in a classical communication
using a shared random binary sequence. 
We investigate a one-time pad and show that 
it provides a secure communication also 
in the context of algorithmic information. 
In Sec. \ref{subsec:Proof},
the main part of the present paper, the security proof of the BB84 
protocol is discussed. 
We introduce a simple BB84 quantum key distribution protocol 
and show that it enables Alice and Bob to share a binary 
sequence that looks almost random to Eve with probability 
exponentially close to $1$.
In Sec. \ref{sec:discussions}, 
we give some discussion of our results and 
future problems. 
\section{
Quantum Kolmogorov Complexity based on Classical Description }
\label{sec:QKC}
Recently some quantum versions of Kolmogorov 
complexity were proposed by a several researchers.   
Svozil \cite{Svozil}, in his pioneering work, 
defined the quantum Kolmogorov complexity 
as 
the minimum classical description length of a quantum state
through a quantum Turing machine \cite{Deutsch,Bernstein}. 
As is easily seen by comparing the cardinality of a set of all the 
programs with that of a set of all the quantum states, 
the value often becomes infinity. 
Vit\'anyi's definition \cite{Vitanyi}, 
while similar to Svozil's, does not 
have this disadvantage. 
He added a term that compensates for the
a difference between a target state and an output state. 
Berthiaume, van Dam, and Laplante \cite{vanDam}
defined 
their quantum Kolmogorov complexity as the length of the shortest 
quantum program that outputs a target state.
The definition was settled and its properties 
were extensively investigated by M\"uller \cite{Muller, MullerPhD}.
Gacs \cite{Gacs} employed a different 
starting point related to the algorithmic probability
to define his quantum Kolmogorov complexity.     
\par
In this paper we employ the definition given by 
Vit\'anyi \cite{Vitanyi}.
The use of Vit\'anyi's definition is justified for the following reason. 
Since, as will be seen in the next section, we are interested in 
the randomness of a classical final key for Eve, to consider its classical 
description is sufficient even if Eve has a quantum state. 
This way of thinking is natural in quantum-information theory.
That is, when one is interested in classical outputs, the inputs to be 
considered are also classical. 
Vit\'anyi gave a description of a one-way quantum Turing machine and 
utilized it to define a prefix quantum Kolmogorov complexity. 
A one-way quantum Turing machine consists of 
four tapes and an internal control. 
(See \cite{Vitanyi} for more details.)
Each tape is a one-way infinite qubit chain
and has a corresponding head on it. 
One of the tapes works as the input tape and 
is read-only from left-to-right.
A program is given on this tape
as an initial condition.
The second tape works as the work tape.
The work tape is initially set to be $0$ for 
all the cells. The head on it can read and write 
a cell and can move in both directions.
The third tape is called an auxiliary tape.
One can put an additional input on this tape. 
The additional input is written to the leftmost qubits and 
can be a quantum state or a classical state. 
This input is needed when one treats conditional 
Kolmogorov complexity. 
The fourth tape works as the output tape. 
It is assumed that after halting
 the state of this tape will not be changed.
The internal control is a quantum system 
described by a finite-dimensional 
Hilbert space 
which has two special orthogonal vectors
$|q_0\rangle$ (initial state) and $|q_f\rangle$ (halting state). 
After each step one makes a measurement of a coarse-grained 
observable $\{|q_f\rangle \langle q_f|,
{\bf 1} -|q_f\rangle \langle q_f|\}$ on the internal control 
to know if 
the computation halts. 
Although there are subtle problems 
\cite{Myers, Ozawa, Popescu, Miya}
in the halting process of 
the quantum Turing machine, we do not get into this problem and 
employ a simple definition of the halting. 
A computation halts at time $t$ if and only  
if 
the probability to observe $q_f$ at time $t$ is one, 
and at any time $t'<t$ a probability to observe 
$q_f$ is zero.  
By using this one-way quantum Turing machine,  
Vit\'anyi defined the quantum Kolmogorov complexity as follows.
He treated 
the length of the shortest classical description of a quantum state. 
That is, the programs of the quantum Turing machine are restricted to 
classical ones.
While the programs must be classical,
the auxiliary inputs can be quantum states. 
We write $U(p,y)=|x\rangle$ if and only if a quantum Turing machine $U$ with 
a classical program $p$ and an auxiliary 
(classical or quantum) input $y$ halts 
and outputs $|x\rangle$. 
The following is the precise description of 
Vit\'anyi's definition.
\begin{definition}\cite{Vitanyi}
The (self-delimiting) quantum Kolmogorov 
complexity of a pure state $|x\rangle$ 
with respect to a one-way quantum Turing machine $U$ with $y$ 
(possibly a quantum state) as conditional input 
given for free is 
\begin{eqnarray*}
K_U(|x\rangle,|\ y)
:=\min_{p,|z\rangle} \{l(p)+ \lceil -\log_2 |\langle z|x\rangle |^2 \rceil :
U(p,y)=|z\rangle\},
\end{eqnarray*}
where $l(p)$ is the length of a classical program $p$, and 
$\lceil a \rceil$ is the smallest integer larger than $a$. 
\end{definition}
The one-way quality of the quantum Turing machine ensures that 
the halting programs compose a prefix-free set. 
Because of this, the length $l(p)$ is defined consistently. 
The term $\lceil -\log_2 |\langle z|x\rangle |^2 \rceil$
represents how insufficiently an output $|z\rangle$ approximates 
the desired output $|x\rangle$. 
This additional term has 
a natural interpretation using the Shannon-Fano code. 
Vit\'anyi has shown the following invariance theorem, 
which is very important.  
\begin{theorem}\cite{Vitanyi}
There is a universal quantum Turing 
machine $U$, such that for all machines $Q$
there is a constant $c_Q$, such that for all quantum states $|x\rangle$
and all auxiliary inputs $y$ we have
\begin{eqnarray*}
K_U(|x\rangle  |\ y)
\leq K_Q(|x\rangle |\ y)
+c_Q.
\end{eqnarray*}
\end{theorem}
Thus the value of 
the quantum Kolmogorov complexity does not depend on 
the choice of the quantum Turing machine if 
one neglects the unimportant constant term $c_Q$. 
Thanks to this theorem, 
one often writes $K$ instead of 
$K_U$.  
Moreover, the following theorem is crucial for our discussion.
\begin{theorem}\label{th:classical}\cite{Vitanyi}
On classical objects (that is, finite binary strings that are all directly 
computable) the quantum Kolmogorov complexity 
coincides up to a fixed additional constant with the self-delimiting 
Kolmogorov complexity. That is, 
there exists a constant $c$ such that for 
any classical binary sequence $|x\rangle$,
\begin{eqnarray*}
\min_q \{l(q): U(q,y)=|x\rangle \}\geq
 K(|x\rangle|\ y)\geq 
\min_q \{l(q): U(q,y)=|x\rangle\} -c
\end{eqnarray*}
holds.
\end{theorem}
According to this theorem, for classical objects 
it essentially suffices to treat only programs that exactly 
output the object.
\section{
Security Proof of Quantum Key Distribution in the light of 
Quantum Algorithmic Information}\label{sec:QKD}
\subsection{Security of one-time pad}\label{subsec:OTP}
The goal of the quantum key distribution is 
to distribute a secret key only between legitimate users. 
In the context of algorithmic information, 
a secret key is nothing but a binary sequence
that looks random to Eve. 
We first show that a random binary sequence shared only
by Alice and Bob does work for secure communication thereafter.
Suppose that Alice and Bob share a common binary sequence $k \in \{0,1\}^M$.
Eve does not know the sequence 
except for its length.
That is, the uncertainty of $k$ for Eve is 
$K(k|M)$. 
 Suppose that Alice sends a message $x\in \{0,1\}^M$ to Bob 
 by one-time pad. That is, Alice sends a 
 binary sequence $x\oplus k \in \{0,1\}^M$ which is known
 also by Eve. 
 Bob, who knows $k$, can decode it easily to obtain 
 $x$. 
In addition, if $k$ is not random, its short description enables 
Eve to reproduce $x$ from any given $x\oplus k$. 
Moreover, we can show the following. 
\begin{theorem}\label{onetimeth}
There exists a constant $c$ (that depends 
only on a choice of a quantum Turing machine) 
such that the following statement holds.
Let $M$ be an arbitrary positive integer and 
let $k\in \{0,1\}^M$ be a binary sequence.
For any $\delta >0$, we define a set $B_{\delta} \subset \{0,1\}^M$ as
\begin{eqnarray*}
B_{\delta}:=\{x|K(x|x\oplus k, M)\leq 
K(k|M) -\delta M -c\}.
\end{eqnarray*}
The size of this $B_{\delta}$ is bounded by
\begin{eqnarray*}
|B_{\delta}|\leq 2^{(1-\delta)M}.
\end{eqnarray*}
\end{theorem}
{\bf Proof:} See Appendix \ref{pfonetime}. 
\par
The following corollary is obvious.
\begin{corollary}
There exists a constant $c$ 
such that the following statement holds.
Let $M$ be an arbitrary positive integer and 
let $k\in \{0,1\}^M$ be a binary sequence
that looks random to Eve, who knows its length only.
That is, $K(k|M)\geq M$ holds. 
For any $\delta >0$, we define a set $B_{\delta} \subset \{0,1\}^M$ as
\begin{eqnarray*}
B_{\delta}:=\{x|K(x|x\oplus k, M)\leq 
(1-\delta) M -c\}.
\end{eqnarray*}
The size of this $B_{\delta}$ is bounded by
\begin{eqnarray*}
|B_{\delta}|\leq 2^{(1-\delta)M}.
\end{eqnarray*}
\end{corollary}
The size $|B_{\delta}|$ in this corollary is thus 
much smaller than $|\{0,1\}^M|=2^M$.
This corollary shows that, 
if Alice and Bob share a random binary sequence
only between them, they can achieve a secret 
communication by one-time pad. 
\par
Let us note a remark. 
One may wonder whether one can show that the size of a set
$\{x|K(x|x\oplus k, M)\leq M-c\}$ is
exponentially small compared with $|\{0,1\}^M|=2^M$. 
It is not possible 
because many $x$'s have a small Kolmogorov complexity 
even if Eve does not know $x\oplus k$. 
For instance, the Kolmogorov 
 complexity of 
 $x=00\ldots 00 \in \{0,1\}^M$ is almost vanishing. 
Thus even $|\{x|K(x|M) \leq M-c\}|$ can be 
comparable with $2^M$, while 
$|\{x|K(x|M) \leq (1-\delta)M -c\}| \leq 2^{(1-\delta)M}
$ holds for $c\geq 0$. 
\subsection{BB84 protocol}\label{subsec:BB84}
As was discussed in the last section, 
if Alice and Bob share a binary sequence 
that is random for Eve, they can communicate 
securely by using the sequence. 
Our goal in the following is to show that  
a quantum key distribution indeed achieves 
this distribution of a random binary sequence. 
In this section, a concrete protocol to be analyzed is 
introduced. 
We consider a quantum key distribution protocol that 
uses a preshared secret key for error correction and
uses 
a public linear code for privacy amplification. 
Although there are more sophisticated or realistic ones,
we treat one of the simplest protocols since our aim is 
to present a different viewpoint from the algorithmic information. 
Let us introduce the protocol. 
\begin{itemize}
\item[(i)]
Alice encodes a probabilistically 
\cite{prob} 
chosen $2N$ bit classical sequence to 
a quantum state of $2N$-qubits with respect to a probabilistically 
\cite{prob} chosen 
basis $b \in \{+,\times\}^{2N}$. 
\item[(ii)] 
After confirming Bob's receipt of all the sent qubits,
Alice announces the basis $b$. 
Bob makes a measurement with the basis $b$ on his qubits to obtain an outcome. 
\item[(iii)] 
Alice probabilistically \cite{prob} chooses half the 
number of $2N$ bits
$T\subset \{1,2,\ldots,2N\}$ ($|T|=N$), 
which are called test bits. The remaining bits 
$I:=\{1,2,\ldots,2N\}\setminus T$ are called information bits.
Alice announces $T$ and 
the classical sequence $z_T\in \{0,1\}^N$
which was encoded to the test bits. 
\item[(iv)]
Alice and Bob check the error rate in the test bits by 
public discussions. If the error rate is larger than 
a preagreed threshold $p$, they abort the protocol. 
\item[(v)]
Alice and Bob perform an error correction by 
the one-time pad using a
preshared secret key. They consume $Nh(p)+const$ secret bits
for this procedure. 
\item[(vi)]
Alice and Bob perform a privacy amplification. 
(See below for the details.)
\end{itemize}
After error correction, Alice and Bob have
a common sifted key $x\in \{0,1\}^N$ (information bits). 
On the other hand, Eve has a quantum state 
that may be
correlated with $x$. 
Due to this correlation, Eve may have a part of the information 
on $x$. Alice and Bob, therefore, cannot use $x$ itself 
as the final key. 
Privacy amplification is a protocol that extracts a shorter final key 
which cannot be guessed by Eve at all.  
The privacy amplification in our protocol 
is performed by use of a linear code.
All players including Eve know a set of 
linear independent vectors
$\{v_1,v_2,\ldots,v_M\} \subset \{0,1\}^N$ which 
span a linear code ${\cal C}$. 
The vectors could be announced before the whole protocol.
Its Hamming distance $d({\cal C})
=\min\{|v|:\ v\neq 0, v\in {\cal C}\}$
is assumed to satisfy $d({\cal C})>2N(p+\epsilon)$, 
where $p$ is the allowed error rate in test bits 
and $\epsilon>0$ is a small security parameter.
The final key is obtained from the sifted key 
by a function $f :\{0,1\}^N \to 
\{0,1\}^M$ which is defined as
\begin{eqnarray*}
f(x)=x\cdot v:=(x\cdot v_1, x\cdot v_2, \ldots , x\cdot v_M).
\end{eqnarray*}
Eve's purpose is to obtain knowledge of $f(x)$.
\subsection{Security proof}\label{subsec:Proof}
Suppose that Alice has chosen a basis $b\in \{0,1\}^{2N}$,
 test bits $T$, 
a value of the test bits $z_T\in\{0,1\}^N$, 
and Bob has obtained $z'_T\in \{0,1\}^N$ as the value of 
the test bits. After (v) in the above protocol
 Eve also knows all of them. 
As is well-known, one can view the protocol also from 
an Ekert 1991 protocol (E91) like setting. 
In the E91 like setting, after the error correction 
there is an entangled state over Alice's information bits, 
Bob's information bits and Eve's apparatus. 
We denote
the state as
$\rho_{b,T,z_T,z'_T}$.
Alice makes a measurement $X_A=\{|x\rangle \langle x|\}$ 
on her information bits to obtain a sifted key $x\in \{0,1\}^N$. 
This measurement changes the state on Bob's information bits and 
Eve's apparatus \cite{aposteriori}. 
We denote the a posteriori state on Bob's information bit and 
Eve's apparatus as $\rho_{x,b,T,z_T,z'_T}$.
We further write its restriction on Eve's apparatus as 
$\rho^E_{x,b,T,z_T,z'_T}$.  
Eve's purpose is to extract information on the final key $f(x)$ 
from this quantum state and her knowledge, $b,T,z_T,z'_T$, and 
$f$.  
Therefore in the context of quantum 
Kolmogorov complexity, 
Eve's uncertainty on the final key is written as 
$K(f(x)|\rho^E_{x,b_T,z_T,z'_T},f,b,T,z_T,z'_T)$
\cite{identify}. 
We prove the following theorem.
\begin{theorem}
There exists a constant $c$ 
(that depends only on the choice of the quantum Turing machine)
such that the following 
statement holds.
For any $N$, any $p$, any $\epsilon >0$, 
any independent vectors $\{v_1,v_2,\ldots,v_M\}$ whose span 
${\cal C}$ satisfies $d({\cal C})> 2N(p+\epsilon)$, and
any $\delta >0$, 
\begin{eqnarray*}
\mbox{Pr}\left(
K(f(x)|\rho^E_{x,b,T,z_T,z'_T},f,b,T,z_T,z'_T)
\leq
M-\delta N -c
\wedge
|z_T \oplus z_T'|<Np
\right)
\leq 2^{-\delta N} +3 e^{-\frac{\epsilon^2}{4}N}
\end{eqnarray*}
holds. 
\end{theorem}
{\bf Proof:}
\\
We
fix a universal quantum Turing machine $U$ and discuss 
the values of the quantum Kolmogorov complexity with respect to it. 
Since $f(x)$ is classical, to discuss 
the quantum Kolmogorov complexity of $f(x)$ it 
essentially suffices to consider 
programs that exactly 
output $f(x)$ thanks to Theorem \ref{th:classical}. 
For each output $x \in \{0,1\}^M$, there is a shortest program
$t_{x,b,T,z_T,z'_T}$ (take an arbitrary one if 
it is not unique) that produces $f(x)$ exactly as 
its output with auxiliary inputs
$\rho^E_{x,b,T,z_T,z'_T}$ and $f,b,T,z_T,z'_T$. 
Although the $t_{x,b,T,z_T,z'_T}$'s may have
different halting times, thanks to a lemma 
proved by M\"uller (Lemma 2.3.4. in \cite{MullerPhD}),
there exists a completely positive map (CP map)
$\Gamma_U: \Sigma({\cal H}_A \otimes {\cal H}_I)
\to \Sigma({\cal H}_O)$
satisfying 
\begin{eqnarray*}
\Gamma_U(\rho^E_{x,b,T,z_T,z'_T}
 \otimes |t_{x,b,T,z_T,z'_T} \rangle \langle 
 t_{x,b,T,z_T,z'_T}|)=|f(x)\rangle \langle f(x)|,
\end{eqnarray*}
where ${\cal H}_A$ is the
Hilbert space for the auxiliary input and ${\cal H}_I$ is 
the Hilbert space for 
programs and ${\cal H}_O =\otimes^M {\bf C}_2$
is the Hilbert space 
for outputs, and $\Sigma({\cal H})$ denotes the set of all the 
density operators on ${\cal H}$. 
\par
For a while we 
proceed with our analysis for fixed $b,T,z_T,z'_T$.
For each $t \in \{0,1\}^*$ (a set of all the finite length 
binary sequences), let 
us define a set ${\cal E}_t^{b,T,z_T,z'_T} \subset \{0,1\}^N$ as 
$
{\cal E}_t^{b,T,z_T,z'_T}=\{x|\ t_{x,b,T,z_T,z'_T}=t \}. 
$
That is, for each $x\in {\cal E}_t^{b,T,z_T,z'_T}$ the program $t$
with auxiliary inputs $\rho^E_{x,b,T,z_T,z'_T}$ and 
$f,b,T,z_T,z'_T$ produces 
exactly $f(x)$.
The set is further decomposed with respect to their outputs as 
${\cal E}_t^{b,T,z_T,z'_T} =\cup_y {\cal E}_t^{b,T,z_T,z'_T}(y)$, 
where ${\cal E}_t^{b,T,z_T,z'_T}(y):=\{x|\ t_{x,b,T,z_T,z'_T}=t, 
f(x)=y\}$.
That is, for each $x\in {\cal E}_t^{b,T,z_T,z'_T}(y)$ 
the program $t$ with
an auxiliary input $\rho^E_{x,b,T,z_T,z'_T}, f,b,T,z_T,z'_T$ 
produces $y$. 
Since the CP map $\Gamma_U$ does not 
increase distinguishability among states, 
for any $x \in {\cal E}_t^{b,T,z_T,z'_T}(y)$ and 
$x' \in {\cal E}_t^{b,T,z_T,z'_T}(y')$
with $y \neq y'$,
$\rho^E_{x,b,T,z_T,z'_T}$ and 
$\rho^E_{x',b,T,z_T,z'_T}$ must be completely distinguishable. 
We denote by $E^{b,T,z_T,z'_T}_t:=\{E^{b,T,z_T,z'_T}_t(y)\}_y$ 
a projection valued measure (PVM) that perfectly
distinguishes states which belong to different $y$. 
That is, 
\begin{eqnarray}
\mbox{tr}\left(E^{b,T,z_T,z'_T}_t(y) \rho^E_{x,b,T,z_T,z'_T}
\right)
=\delta_{f(x) y}
\label{kubetu}
\end{eqnarray}
holds for each $x$ and $y$. 
\par
Let us consider the problem in an E91 like setting.
Now Alice, Bob, and Eve have a state $\rho_{b,T,z_T,z'_T}$ 
over their systems. 
 For an arbitrary fixed finite $L \subset \{0,1\}^*$, let us consider 
 an observable over Alice's information bits and Eve's apparatus; 
 \begin{eqnarray*}
 Q^{b,T,z_T,z'_T}_L:=
 \sum_{t\in L}\sum_y A^{b,T,z_T,z'_T}_t(y)\otimes 
 E^{b,T,z_T,z'_T}_t(y),
 \end{eqnarray*}
 where $A^{b,T,z_T,z'_T}_t(y)$ is defined as
 \begin{eqnarray*}
 A^{b,T,z_T,z'_T}_t(y):=\sum_{x\in {\cal E}^{b,T,z_T,z'_T}_t(y)}
  |x\rangle \langle x|.
 \end{eqnarray*}
One can easily show that this $Q^{b,T,z_T,z'_T}_L$ 
is a projection operator.
We hereafter consider an
 expectation value of this projection operator with respect to 
the state $\rho_{b,T,z_T,z'_T}$. 
($Q^{b,T,z_T,z'_T}_L$ is naturally 
identified with an operator
$Q^{b,T,z_T,z'_T}_L \otimes {\bf 1}_B$ 
on Alice, Bob, and Eve's tripartite system.) 
One can write it as follows:
\begin{eqnarray*}
\mbox{tr}(\rho_{b,T,z_T,z'_T}Q^{b,T,z_T,z'_T}_L)
=
\langle Q^{b,T,z_T,z'_T}_L \rangle_{b,T,z_T,z'_T}
\\
=\sum_{t\in L} 
\sum_y \langle A^{b,T,z_T,z'_T}_t(y)\otimes E^{b,T,z_T,z'_T}_t(y)
\rangle_{b,T,z_T,z'_T}
\\
=\sum_{t\in L} \sum_y 
\sum_{x\in {\cal E}^{b,T,z_T,z'_T}_t(y)}
\langle |x\rangle \langle x|\otimes E^{b,T,z_T,z'_T}_t(y)
\rangle_{b,T,z_T,z'_T},
\end{eqnarray*}
where we put  
$\langle \ \cdot \ \rangle_{b,T,z_T,z'_T}
=\mbox{tr}(\rho_{b,T,z_T,z'_T}\ \cdot\ )$. 
If we consider Alice's measurement $X_A=\{|x\rangle\langle x|\}$
on her information bits and denote by
$p(x|b,T,z_T,z'_T)$ the probability to obtain $x$, 
it is represented as
\begin{eqnarray}
\sum_{t\in L} \sum_y 
\sum_{x\in {\cal E}^{b,T,z_T,z'_T}_t(y)}
\langle |x\rangle \langle x|\otimes E^{b,T,z_T,z'_T}_t(y)
\rangle_{b,T,z_T,z'_T}
\nonumber
\\
=\sum_{t\in L} \sum_y 
\sum_{x\in {\cal E}^{b,T,z_T,z'_T}_t(y)}
p(x|b,T,z_T,z'_T)\mbox{tr}(\rho^E_{x,b,T,z_T,z'_T} E^{b,T,z_T,z'_T}_t
(y))
\nonumber
\\
=\sum_{t\in L} \sum_y 
\sum_{x\in {\cal E}^{b,T,z_T,z'_T}_t(y)}p(x|b,T,z_T,z'_T)
=\mbox{Pr}\left(
x\in \bigcup_{t\in L}{\cal E}^{b,T,z_T,z'_T}_t|\
b,T,z_T,z'_T\right),
\label{eq5}
\end{eqnarray}
where we have used the condition (\ref{kubetu}). 
\par
In addition, this quantity can be represented in 
a different form (see Appendix \ref{pflemma} for its proof). 
\begin{lemma}\label{lemmada}
Suppose that Alice
virtually makes a measurement on her 
information bits with a PVM
$Z_A:=\{|\overline{z}\rangle \langle \overline{z}|\}$ 
which is conjugate to $X_A=\{|x\rangle \langle x|\}$ 
that is actually measured to obtain a sifted key,
and Bob virtually makes a measurement on his information bits
with $Z_B:=\{|\overline{z}\rangle \langle \overline{z}|\}$
which is conjugate to $X_B$ that is actually measured. 
We denote their outcomes $z_I$ and $z'_I$. 
It holds that
\begin{eqnarray}
\mbox{tr}(\rho_{b,T,z_T,z'_T}Q^{b,T,z_T,z'_T}_L)
\leq |L| 2^{-M}
+3\sqrt{\mbox{Pr}(|z_I \oplus z'_I|>N(p+\epsilon)|b,T,z_T,z'_T)},
\label{eq4}
\end{eqnarray}
where the second term in the 
right hand side is the square root of the probability to 
obtain distant $z_I$ and $z'_I$ with respect to 
a state $\rho_{b,T,z_T,z'_T}$.  
\end{lemma}
Combining these different expressions (\ref{eq5}) and 
(\ref{eq4}),
we obtain
\begin{eqnarray*}
\mbox{Pr}\left(
x\in \bigcup_{t\in L}{\cal E}^{b,T,z_T,z'_T}_t|\
b,T,z_T,z'_T\right)
\leq 
|L| 2^{-M}
+3 \sqrt{
\mbox{Pr}\left(|z_I\oplus z'_I|> N(p+\epsilon)|\ b,T,z_T,z'_T
\right)
}.
\end{eqnarray*}
Now if $L$ is taken as $L:=\{t|\ l(t)\leq M-\delta N\}$, 
since $|L|\leq 2^{M-\delta N}$ holds
the above inequality can be rewritten as 
\begin{eqnarray*}
\mbox{Pr}\left(x\in \bigcup_{t: l(t) \leq M-\delta N
}{\cal E}^{b,T,z_T,z'_T}_t|\
b,T,z_T,z'_T
\right)
\leq 
2^{-\delta N}
+3 \sqrt{
\mbox{Pr}\left(
|z_I\oplus z'_I|> N(p+\epsilon)|\ b,T,z_T,z'_T
\right)
}.
\end{eqnarray*}
Thanks to Theorem \ref{th:classical}, 
there exists a constant $c$ such that,
if $x$ satisfies 
$K(f(x)| \rho^E_{x,b,T,z_T,z'_T},b,T,z_T,z'_T,f)
\leq M-\delta N -c $, then
$l(t_{x,b,T,z_T,z'_T})
\leq M-\delta N$ follows. 
That is, we obtain 
\begin{eqnarray*}
\mbox{Pr}\left(
K(f(x)| \rho^E_{x,b,T,z_T,z'_T},b,T,z_T,z'_T,f)
\leq M-\delta N -c
|\
b,T,z_T,z'_T
\right)
\leq 
2^{-\delta N}
\\
+3 \sqrt{
\mbox{Pr}\left(
|z_I\oplus z'_I|> N(p+\epsilon)|\ b,T,z_T,z'_T
\right)
}.
\end{eqnarray*}
We multiply both sides of this inequality by 
$p(b,T,z_T,z'_T)$ which is defined as the
probability to obtain $b, T, z_T, z'_T$ 
and take a summation with respect to 
$b,T,z_T,z'_T$ for all $b,T$, and $z_T,z'_T$ with $|z_T \oplus 
z'_T|
\leq Np$, and use Jensen's inequality. We finally derive
\begin{eqnarray*}
\mbox{Pr}\left(K(f(x)|\rho^E_{x,b,T,z_T,z'_T},b,T,z_T,z'_T,f)
\leq M- \delta N -c
\wedge |z_T \oplus z'_T|<Np
\right)
\leq \mbox{Pr}\left(|z_T \oplus z'_T|<Np
\right) 
2^{-\delta N}
\\
 +
3 
\sqrt{\mbox{Pr}\left(|z_T \oplus z'_T|\leq Np\right)}
\sqrt{\mbox{Pr}\left(
|z_I \oplus z'_I| > N(p+\epsilon), |z_T \oplus z'_T|\leq Np
\right)}.
\end{eqnarray*}
The second term of the right hand side is 
bounded by Hoeffding's lemma 
as $\mbox{Pr}(|z_I \oplus z'_I|> N(p+\epsilon),
|z_T\oplus z'_T|\leq Np) \leq e^{-\frac{\epsilon^2}{2}N}$
(see e.g. \cite{Boykin}). 
We thus obtain 
\begin{eqnarray*}
\mbox{Pr}\left(
K(f(x)|\rho^E_{x,b,T,z_T,z'_T},b,T,z_T,z'_T,f)\leq
M-\delta N -c
\wedge |z_T \oplus z'_T|<Np
\right)
\leq 
2^{-\delta N}
+
3 e^{-\frac{\epsilon^2}{4}N} .
\end{eqnarray*}
This ends the proof.
\hfill Q.E.D.
\section{discussions}\label{sec:discussions}
In this paper, we considered the security of the
quantum key distribution 
protocol in the light of quantum algorithmic information. 
We employed the quantum Kolmogorov complexity defined 
by 
Vit\'anyi as the fundamental quantity, discussed a possible 
security criterion, and showed that the simple BB84 protocol 
satisfies it. According to the main theorem, 
a probability for Eve to obtain an almost random 
final key is exponentially close to $1$. 
The length of the final keys $M$ is determined by 
a condition for the Hamming distance. One can take 
it as $M \simeq N(1-h(2(p +\epsilon)))$. Since 
the legitimate users have consumed $Nh(p+\epsilon)$ bits 
for the error correction, the length of the 
key produced amounts to 
$N(1-h(2(p+\epsilon))-h(p+\epsilon))$. It coincides with 
the rate obtained in \cite{Boykin} where the security criterion 
was based on Shannon's information theory.
\par
Although we hope that the present work can be a 
first step toward the study of quantum 
cryptography from the viewpoint of 
quantum algorithmic information,  
there still remain a lot of things to be investigated.
The security criterion employed in this paper utilizes 
the quantum Kolmogorov complexity, but it still needs 
the probability. Therefore, 
the original motivation of the algorithmic information 
theory, in some sense, 
has not been perfectly accomplished. 
Comparison between security notions based on algorithmic 
information and Shannon's information is an important 
future problem to be considered. 
While the simple BB84 protocol satisfies 
both criteria, it is not clear whether 
one can be derived from another in some sense. 
The relation between these criteria 
will become more subtle if we will deepen our algorithmic information 
theoretical discussion so as to avoid an appearance of probability completely. 
For instance, as was shown, in the one-time pad protocol, 
while an individual secret key cannot be discussed 
in the conventional Shannon's information theory, it can be treated in the algorithmic information theory. 
In addition, as we noted in Sec. \ref{sec:QKC}, 
there are some other definitions of quantum Kolmogorov 
complexity. It is interesting to investigate 
whether one can apply them to the security problem. 
Application of our argument to other protocols 
will be another interesting problem. 
\begin{acknowledgements}
The authors thank 
K. Imafuku, K. Nuida, and an anonymous referee 
for their helpful comments. 
\end{acknowledgements}

\appendix

\section{Technical lemmas}
\begin{lemma}\label{lemma1}.
For any state $\rho$ and
any projection operators $Q$ and $P$, 
it holds that
\begin{eqnarray*}
\left| \mbox{tr}(\rho Q)-\mbox{tr}(PQP)\right|
\leq 3 \mbox{tr}(\rho ({\bf 1}-P))^{1/2}.
\end{eqnarray*} 
\end{lemma}
{\bf Proof:}
Since ${\bf 1}=P+({\bf 1}-P)$ holds, 
$Q$ can be decomposed as
\begin{eqnarray*}
Q={\bf 1}Q{\bf 1}=PQP+PQ({\bf 1}-P) +({\bf 1}-P)QP
+({\bf 1}-P)Q({\bf 1}-P).
\end{eqnarray*} 
Thus we obtain
\begin{eqnarray*}
\left| \mbox{tr}(\rho Q)-
\mbox{tr}(\rho PQP) \right|
\leq |\mbox{tr}(\rho PQ({\bf 1}-P))|
+|\mbox{tr}(\rho ({\bf 1}-P)QP)|
+|\mbox{tr}(\rho ({\bf 1}-P)Q({\bf 1}-P))|.
\end{eqnarray*}
The Cauchy-Schwarz inequality bounds the first term in the right-hand side
as
\begin{eqnarray*}
|\mbox{tr}(\rho PQ({\bf 1}-P))|
=\mbox{tr}(\rho PQP)^{1/2} \mbox{tr}(\rho ({\bf 1}-P))^{1/2}
\leq \mbox{tr}(\rho({\bf 1}-P))^{1/2}.
\end{eqnarray*}
Other terms can be bounded in a similar manner. 
This ends the proof. 
\hfill Q.E.D.
\begin{lemma}\label{lemma2}
For given linearly independent vectors $\{v_1,v_2,\ldots,v_M\} 
\subset \{0,1\}^N$, we define $f:\{0,1\}^N \to \{0,1\}^M$ as
$f(x)=(x\cdot v_1,x\cdot v_2,\ldots,x\cdot v_M)$. 
Let ${\cal C}$ be a code generated by $\{v_1,v_2,\ldots,v_M\}$
and $d({\cal C})$ be its Hamming distance.
For $s,t \in \{0,1\}^N$ satisfying 
$|s|, |t|< \frac{d({\cal C})}{2} $ and for any $y \in \{0,1\}^M$, 
\begin{eqnarray*}
\sum_{x:f(x)=y}(-1)^{x\cdot (s\oplus t)}
=\delta_{st}2^{N-M}
\end{eqnarray*}
holds, where $\delta_{st}$ is Kronecker's delta. 
\end{lemma}
{\bf Proof:}
If we fix an element $w_y \in \{0,1\}^N$ satisfying 
$f(w_y)=y$, $\{x|f(x)=y\}$ is represented as 
$w_y \oplus {\cal C}^{\perp}$. 
Thus we obtain 
\begin{eqnarray*}
\sum_{x:f(x)=y}(-1)^{x\cdot (s\oplus t)}
=(-1)^{w_y \cdot (s\oplus t)}
\sum_{x\in {\cal C}^{\perp}}
(-1)^{x\cdot (s\oplus t)}.
\end{eqnarray*}
For $s\oplus t \in {\cal C}$, it gives 
$2^{N-M}(-1)^{w_y \cdot (s\oplus t)}$. 
Since $|s \oplus t|\leq |s|+|t|< d({\cal C})$ holds, 
$s\oplus t\in {\cal C}$ means $s=t$. 
For $s\oplus t \notin {\cal C}$,
thanks to Lemma D.1 in \cite{Boykin}, 
\begin{eqnarray*}
\sum_{x\in {\cal C}^{\perp}}(-1)^{x\cdot (s\oplus t)}
=0 
\end{eqnarray*}
holds.
This ends the proof.
\hfill Q.E.D.
\section{Proof of Theorem \ref{onetimeth}}\label{pfonetime}
{\bf Proof of Theorem \ref{onetimeth}}
According to the fundamental properties 
\cite{LiVitanyi} of Kolmogorov complexity 
it is known that
\begin{eqnarray*}
|K(x,k|M)- (K(k|M)+K(x|k,K(k),M))|\leq c_1
\end{eqnarray*}
holds for some constant $c_1$. 
(The proof also holds 
for the quantum Kolmogorov complexity thanks to Theorem \ref{th:classical}.)
For a fixed $\delta >0$, we define 
a set ${\cal D}_{\delta} \subset \{0,1\}^M$ as 
\begin{eqnarray*}
{\cal D}_{\delta}:=\{x|K(x|k,K(k),M)\leq (1-\delta) M\}. 
\end{eqnarray*}
It can be easily shown that 
$|{\cal D}_{\delta}|\leq 2^{(1-\delta)M}$ holds. 
Now let us consider its complement 
${\cal D}_{\delta}^{c}
=\{x\in \{0,1\}^M|
K(x|k,K(k),M)> (1-\delta)M\}$. 
For $x\in {\cal D}_{\delta}^c$, $K(x,k|M)> K(k|M)+
(1-\delta)M -c_1$ holds. 
By the way we have, in general, 
\begin{eqnarray*}
K(x,k|M) =K(x\oplus k,k|M)+c_2
\leq K(x\oplus k|M)+K(x|x\oplus k,M) +c_3
\end{eqnarray*}
for some $c_2,c_3$.
Thus, for $x\in {\cal D}_{\delta}^c$, we have 
\begin{eqnarray*}
K(x\oplus k|M) + K(x|x\oplus k,M)+c_3
> K(k|M)+(1-\delta)M - c_1.
\end{eqnarray*}
Since $K(x\oplus k|M)\leq M +c_4$ holds for some $c_4$,
if we put $c=c_1+c_3+c_4$ we obtain
\begin{eqnarray*}
K(x|x\oplus k,M)>
K(k|M)-\delta M -c
\end{eqnarray*}
for $x \in {\cal D}^c_{\delta}$. 
Thus ${\cal D}_{\delta}^c \subset 
B_{\delta}^c$ and 
$B_{\delta} \subset {\cal D}_{\delta}$ holds.
Thanks to $|{\cal D}_{\delta}|\leq 2^{(1-\delta)M}$, 
this ends the proof. \hfill Q.E.D.
\section{Proof of Lemma \ref{lemmada}}\label{pflemma}
{\bf Proof of Lemma \ref{lemmada}}
Let $\rho_{b,T,z_T,z'_T}$ be a state over 
Alice's information bits, Bob's information bits, and 
Eve's apparatus.
Suppose that Bob virtually makes a measurement 
of $Z_B=\{|\overline{z}\rangle \langle \overline{z}|\}$ 
on his system (information bits).
This observable is 
conjugate with $X_B$, which is actually measured 
by Bob to obtain a sifted key.
Suppose that Bob obtains 
an outcome $z'_I$. 
We denote by $p(z'_I|b,T,z_T,z'_T)$ a probability to obtain 
$z'_I$.
The a posteriori state on Alice's information bits and Eve's apparatus 
is denoted as $\overline{\rho^{b,T,z_T,z'_T}_{z'_I}}$. 
\par 
Define a projection operator $P^{b,T}_{z'_I}$ on Alice's information bits 
as 
\begin{eqnarray*}
P^{b,T}_{z'_I}:=\sum_{s}^{|s|\leq N(p+\epsilon)}|\overline{
z'_I\oplus s}\rangle 
\langle \overline{z'_I \oplus s}|.
\end{eqnarray*}
Applying Lemma \ref{lemma1} with $P^{b,T}_{z'_I}=P$,
$Q^{b,T,z_T,z'_T}_L=Q$ and $\overline{\rho^{b,T,z_T,z'_T}_{z'_I}}=\rho$, 
we obtain 
\begin{eqnarray}
|\langle Q^{b,T,z_T,z'_T}_L
\rangle_{b,T,z_T,z'_T,z'_I}
 - \langle P^{b,T}_{z'_I}
Q^{b,T,z_T,z'_T}_L P^{b,T}_{z'_I}\rangle_{b,T,z_T,z'_T,z'_I} |
\leq 3 \langle {\bf 1}-P^{b,T}_{z'_I}\rangle_{b,T,z_T,z'_T,z'_I}^{1/2},
\label{eq1}
\end{eqnarray} 
where we put $\langle \ \cdot\ \rangle_{b,T,z_T,z'_T,z'_I}
=\mbox{tr}(\overline{\rho^{b,T,z_T,z'_T}_{z'_I}}(\ \cdot \ ))$. 
In addition, if we introduce 
$A^{b,T}(y):=\sum_{x}^{f(x)=y}|x\rangle \langle x|$, 
it satisfies $A^{b,T}(y)\geq A^{b,T,z_T,z'_T}_t(y)$. 
Thus one can easily show that 
\begin{eqnarray*}
Q^{b,T,z_T,z'_T}_L
\leq \sum_{t \in L} \sum_y
A^{b,T}(y)\otimes E^{b,T,z_T,z'_T}_t(y) 
\end{eqnarray*}
holds.
It follows that
\begin{eqnarray}
 P^{b,T}_{z'_I}
Q^{b,T,z_T,z'_T}_L P^{b,T}_{z'_I}
\leq 
 \sum_{t \in L} \sum_y
 P^{b,T}_{z'_I}
A^{b,T}(y) P^{b,T}_{z'_I}
\otimes E^{b,T,z_T,z'_T}_t(y).
\label{eq2} 
\end{eqnarray}
Combining Eqs.(\ref{eq1}) and (\ref{eq2}), 
 we obtain the inequality
\begin{eqnarray}
\mbox{tr}(\overline{\rho^{b,T,z_T,z'_T}_{z'_I}}Q^{b,T,z_T,z'_T}_L)
=
\langle Q^{b,T,z_T,z'_T}_L\rangle_{b,T,z_T,z'_T,z'_I}
\nonumber 
\\
\leq
\sum_{t\in L}\sum_y
\langle P^{b,T}_{z'_I}A^{b,T}(y)P^{b,T}_{z'_I} \otimes 
E^{b,T,z_T,z'_T}_t(y)\rangle_{b,T,z_T,z'_T,z'_I} 
+3 \langle {\bf 1}-P^{b,T}_{z'_I}\rangle_{b,T,z_T,z'_T,z'_I}^{1/2}.
\label{eq3}
\end{eqnarray}
$\langle P^{b,T}_{z'_I}A^{b,T}(y)P^{b,T}_{z'_I}
 \otimes E^{b,T,z_T,z'_T}_t(y)\rangle_{b,T,z_T,z'_T,z'_I}$ 
 in the first term on the right-hand side of Eq.(\ref{eq3})
is estimated as follows. 
Suppose that, with respect to $\overline{\rho^{b,T,z_T,z'_T}_{z'_I}}$,
Eve made a measurement of the PVM
$E^{b,T,z_T,z'_T}_t$ and obtained $y$. 
The probability to obtain $y$ is denoted as
$p(y|b,T,z_T,z'_T,z'_I)$. 
The a posteriori state over Alice's information bits is denoted 
as $\overline{\rho^{b,T,z_T,z'_T}_{z'_I,y}}$. 
We write its diagonalization as 
$\overline{\rho^{b,T,z_T,z'_T}_{z'_I,y}}
 =\sum_{\nu} \lambda_{\nu} |\phi_{\nu}\rangle 
\langle \phi_{\nu}|$. 
The vector $|\phi_{\nu}\rangle$ has a expansion 
$|\phi_{\nu}\rangle =\sum_s c^{\nu}_s |\overline{z'_I\oplus s}\rangle$. 
Now $\langle \phi_{\nu}|P^{b,T}_{z'_I}
A^{b,T}(y)P^{b,T}_{z'_I}|\phi_{\nu}\rangle$
is calculated as
\begin{eqnarray*}
\langle \phi_{\nu}|P^{b,T}_{z'_I}
A^{b,T}(y)P^{b,T}_{z'_I}|\phi_{\nu}\rangle
=\sum_{s}^{|s|\leq N(p+\epsilon)}
\sum_{t}^{|t|\leq N(p+\epsilon)}
\overline{c^{\nu}_s} c^{\nu}_t
\langle \overline{z'_I \oplus s}|
A^{b,T}(y) |\overline{z'_I \oplus t}\rangle 
\\
=\sum_{s}^{|s|\leq N(p+\epsilon)}
\sum_{t}^{|t|\leq N(p+\epsilon)}
\overline{c^{\nu}_s} c^{\nu}_t
2^{-N}
\sum_{x}^{f(x)=y}
(-1)^{x\cdot (s\oplus t)}
\leq 2^{-M},
\end{eqnarray*}  
where we have used 
lemma \ref{lemma2} 
and $\sum_s |c^{\nu}_s|^2 =1$
to obtain the last inequality.
 We thus obtain for each $y$ and $z'_I$
 \begin{eqnarray*}
 \frac{ \langle P^{b,T}_{z'_I}
 A^{b,T}(y) P^{b,T}_{z'_I}\otimes E^{b,T,z_T,z'_T}_t(y)
 \rangle_{b,T,z_T,z'_T,z'_I}}{
 p(y|b,T,z_T,z'_T,z'_I)
 }=\langle P^{b,T}_{z_I} A^{b,T}(y)P^{b,T}_{z'_I} 
 \rangle_{b,T,z_T,z'_T,z'_I,y}
 \leq 2^{-M}.
 \end{eqnarray*}
 Multiplying both sides of this inequality with 
 $p(y|b,T,z_T,z'_T,z'_I)$ and 
 summing it up with respect to $y$, 
 we obtain
 \begin{eqnarray*}
 \sum_y \mbox{tr}(\overline{\rho^{b,T,z_T,z'_T}_{z'_I}}(
  P^{b,T}_{z'_I} A^{b,T}(y)
  P^{b,T}_{z'_I} \otimes E^{b,T,z_T,z'_T}_t(y)))
 \leq 2^{-M}. 
 \end{eqnarray*}
Summation of this inequality over $t \in L$ further gives
\begin{eqnarray*}
\sum_{t\in L}\sum_y
\langle P^{b,T}_{z'_I}A^{b,T}(y)P^{b,T}_{z'_I} \otimes 
E^{b,T,z_T,z'_T}_t(y)\rangle_{b,T,z_T,z'_T,z'_I} 
\leq |L|2^{-M}.
\end{eqnarray*}
We next estimate the second term 
$3\langle {\bf 1}-P^{b,T}_{z'_I}\rangle_{b,T,z_T,z'_T,z'_I}^{1/2}$
in Eq.(\ref{eq3}). 
This term can be represented in a simple form by considering 
Alice's measurement on her information bits with 
$Z_A:=\{|\overline{z_I}\rangle \langle \overline{z_I}|\}$
which is conjugate to $X_A=\{|x\rangle\langle x|\}$ that is 
actually measured to obtain a sifted key in the E91 like picture. 
One can show
\begin{eqnarray*}
\langle {\bf 1}-P^{b,T}_{z'_I}\rangle_{b,T,z_T,z'_T,z'_I}
=\mbox{Pr}\left(
|z_I \oplus z'_I| > N(p+\epsilon)|\ b,T,z_T,z'_T,z'_I
\right),
\end{eqnarray*}
where the right-hand side is the probability for Alice to 
obtain a distant $z_I$ from Bob's $z'_I$.
Combining the above estimates, we obtain
\begin{eqnarray*}
\langle Q^{b,T,z_T,z'_T}_L\rangle_{b,T,z_T,z'_T,z'_I}
\leq
|L|2^{-M}
+3 \sqrt{\mbox{Pr}\left(
|z_I \oplus z'_I| > N(p+\epsilon)|\ b,T,z_T,z'_T,z'_I\right)
}.
\end{eqnarray*}
We multiply both sides of this inequality by 
$p(z'_I|b,T,z_T,z'_T)$ 
and take a summation over $z'_I$ to 
obtain
\begin{eqnarray*}
\langle Q^{b,T,z_T,z'_T}_L \rangle_{b,T,z_T,z'_T}
\leq |L| 2^{-M}
+3 \sqrt{
\mbox{Pr}\left(
|z_I\oplus z'_I|> N(p+\epsilon)|\ b,T,z_T,z'_T\right)
},
\end{eqnarray*}
where we have used Jensen's inequality once. 
\hfill Q.E.D.


\begin{thebibliography}{9}
\bibitem{note}
Not all the security notions use the Shannon entropy explicitly. 
They can be interpreted in any case in terms of probability theory.
\bibitem{prob}The word ``probabilistically" in this section is used to 
mean ``randomly in a probabilistic sense." 
That is, 
we use an unbiased probability $1/|\Omega|$ to choose a sample 
from a sample space $\Omega$ (say $\Omega=\{0,1\}^{2N}$).
(To avoid a 
possible confusion of this with randomness in 
the algorithmic sense, we just write ``probabilistically.")
\bibitem{identify}
To treat $\rho^E_{x,b_T,z_T,z'_T}$ as an auxiliary input
for a quantum Turing machine, Eve's apparatus is
identified with a system consisting of qubits. 
Our discussion does not depend on this identification.
\end{thebibliography}
\end{document}